\documentclass[journal]{IEEEtran}

\usepackage{pgfplots}
\usepackage[bookmarks,colorlinks]{hyperref}
\usepackage[linesnumbered,ruled,lined]{algorithm2e}
\usepackage{enumitem}
\usepackage{algpseudocode}
\usetikzlibrary{shapes.multipart,intersections}
\usepackage{cite}
\usepackage{amsmath,amssymb,amsfonts,amsthm,steinmetz}
\usepackage{graphicx}
\usepackage{mathrsfs}  
\usepackage{textcomp}
\usepackage{acronym}
\usepackage{xcolor}
\usepackage{upgreek,xspace}
\usepackage{array}
\usepackage{tikz}
\usetikzlibrary{calc}
\makeatletter
\newcommand{\gettikzxy}[3]{%
  \tikz@scan@one@point\pgfutil@firstofone#1\relax
  \edef#2{\the\pgf@x}%
  \edef#3{\the\pgf@y}%
}
\makeatother

\usepackage[draft]{todonotes}

\usepackage{esvect}

\usepackage{times}
\usepackage{bm}
\usepackage{amsmath}
\usepackage{amssymb}
\usepackage{stmaryrd}
\usepackage{babel}
\usepackage{graphics, graphicx}
\usepackage{xcolor}
\usepackage{gensymb}
\usepackage{cite}
\usepackage{enumitem}
\usepackage{url}

\begin{document}

\title{Beyond-Diagonal Dynamic Metasurface Antenna}

\author{Hugo~Prod'homme~and~Philipp~del~Hougne,~\IEEEmembership{Member,~IEEE}
\thanks{This work was supported in part by the ANR France 2030 program (project ANR-22-PEFT-0005), the ANR PRCI program (project ANR-22-CE93-0010), and CREACH Labs (project AdverPhy).}
\thanks{
H.~Prod'homme and P.~del~Hougne are with Univ Rennes, CNRS, IETR - UMR 6164, F-35000, Rennes, France (e-mail: \{hugo.prodhomme; philipp.del-hougne\}@univ-rennes.fr).
}
\thanks{\textit{(Corresponding Author: Philipp del Hougne.)}}
}

\maketitle

\begin{abstract}
Dynamic metasurface antennas (DMAs) are an emerging technology for next-generation wireless base stations, distinguished by hybrid analog/digital beamforming capabilities with low hardware complexity. 
However, the intrinsic coupling between meta-atoms is fixed by \textit{static} waveguide or cavity structures in existing DMAs, which fundamentally constrains the achievable performance. 
Here, we introduce \textit{reconfigurable} intrinsic coupling mechanisms between meta-atoms, yielding finer control over the DMA's analog signal processing capabilities. This novel hardware is coined ``beyond-diagonal DMA'' (BD-DMA), in line with established BD-RIS terminology. Considering realistic hardware constraints, we derive a physics-consistent system model revealing (correlated) ``beyond-diagonal'' programmability. We also present an equivalent formulation with (uncorrelated) ``diagonal'' programmability. Based on the latter, we propose a general and efficient mutual-coupling-aware optimization algorithm. Physics-consistent simulations validate the performance enhancement enabled by reconfigurable intrinsic coupling mechanisms in BD-DMAs. The BD-DMA benefits grow with the mutual coupling strength.
\end{abstract}

\begin{IEEEkeywords}
Dynamic metasurface antenna, beyond-diagonal reconfigurable intelligent surface, reconfigurable inter-element intrinsic coupling, mutual coupling, coupled-dipole formalism.
\end{IEEEkeywords}

\section{Introduction}

Next-generation base stations are expected to serve massive numbers of users under stringent cost and efficiency requirements. Conventional antenna array technology would rely on countless radiofrequency (RF) chains, potentially implying prohibitive cost, power consumption, weight and footprint. An emerging technology promising to overcome these challenges is the dynamic metasurface antenna (DMA). Recent studies have explored the potential of DMAs in 6G massive MIMO wireless communications~\cite{yoo2018enhancing,shlezinger2019dynamic,DMA_Mag,you2022energy,zhang2022beam}, end-to-end optimized integrated sensing and wave-domain computing~\cite{del2020learned,qian2022noise}, wireless power transfer~\cite{azarbahram2024waveform,zhang2025near}, etc.

A DMA consists of a surface patterned with metamaterial elements. These so-called meta-atoms are coupled to the feeds (and other meta-atoms) via waveguide~\cite{sleasman2015dynamic} or cavity~\cite{sleasman2020implementation} structures. Tunable lumped elements are integrated into the meta-atoms to endow their radiation properties with programmability (e.g., on/off switching). The field radiated by the DMA depends on the tunable meta-atoms' configurations and their excitation via the waveguide or cavity structure. As an inherent by-product of this architecture, DMAs hence realize hybrid analog/digital beamforming without requiring additional dedicated analog combining circuitry~\cite{DMA_Mag}. 

Due to mutual coupling (MC) between meta-atoms, the excitation of a given meta-atom generally depends on the configuration of the other meta-atoms. MC between meta-atoms is to date usually neglected~\cite{smith2017analysis} or mitigated~\cite{boyarsky2021electronically} because it results in a non-linear mapping from configuration to radiation pattern, as seen in physics-consistent DMA models~\cite{yoo2019analytic,yoo2019analytic2,williams2022electromagnetic,prod2025benefits}. Yet, stronger MC between meta-atoms boosts the  radiated field's sensitivity to the DMA configuration, and thereby the ability to optimize a DMA for a desired wireless functionality~\cite{prod2024mutual,prod2025benefits}. This insight highlights the largely untapped potential of DMAs with strongly coupled meta-atoms, such as cavity-backed DMAs \cite{yoo2018enhancing,sleasman2020implementation,prod2024mutual,prod2025benefits}, combined with MC-aware optimizations. 

The coupling between any two meta-atoms is mediated by (i) wave propagation in waveguides or cavities, and (ii) scattering off other meta-atoms. 
While (ii) depends on the configuration of the other meta-atoms, (i) does not  -- (i) is an \textit{intrinsic} property of the DMA hardware devoid of the meta-atoms.
In conventional DMAs, each tunable lumped element is part of a meta-atom because it serves to \textit{locally} program the meta-atom's radiation properties; hence, the hardware devoid of the meta-atoms is static and the intrinsic MC is fixed. 
The fixed nature of the intrinsic MC fundamentally limits the 
physical-layer programmability of conventional DMAs, and consequently also their hybrid analog/digital beamforming, which is always characterized by a diagonal matrix~\cite{prod2025benefits} (see also details in Sec.~\ref{sec_system_model} below).

In this Letter, we explore potential performance gains enabled by \textit{reconfigurable rather than fixed intrinsic MC between the meta-atoms}. To that end, we introduce a novel class of DMA architectures in which additional tunable lumped elements are integrated into the waveguide or cavity structure that couples the feeds and meta-atoms. Thanks to these additional tunable elements, the DMA hardware devoid of the meta-atoms is no longer static. 
Indeed, these additional tunable elements parametrize the intrinsic MC between meta-atoms as opposed to parametrizing the meta-atoms' local properties. 
We coin this novel architecture ``beyond-diagonal DMA'' (BD-DMA), following the established term ``beyond-diagonal reconfigurable intelligent surface'' (BD-RIS)~\cite{shen2021modeling,li2023reconfigurable,li2024beyond} for RIS with tunable inter-element intrinsic couplings.\footnote{The key difference between a BD-DMA [resp. DMA] and a BD-RIS [resp. RIS] is that the former has feeds via which signals are injected into or received from the radio environment whereas the latter merely scatters signals already existing in the radio environment.} 

The term ``beyond-diagonal'' in BD-RIS and BD-DMA refers to the fact that common representations of systems with tunable intrinsic MC feature both diagonal and non-diagonal programmable matrix elements. Theoretical studies on BD-RIS tend to treat these programmable matrix elements as independent from each other. However, their programmability is correlated if practical hardware implementations are considered~\cite{del2025physics,tapie2025beyond}. The truly independently controllable degrees of freedom are always individually tunable lumped elements. For any wave system parametrized by tunable lumped elements there exists a physics-consistent description in which the programmability is captured by a diagonal matrix whose diagonal entries can be controlled independently, as recently pointed out for BD-RIS in~\cite{del2025physics}. The diagonal representation of systems with tunable intrinsic MC is  directly compatible with  algorithms developed for systems with static intrinsic MC. We develop and apply these insights in this Letter to efficiently optimize BD-DMAs.

Our contributions are summarized as follows. \textit{First}, we derive in Sec.~\ref{sec_system_model} a physics-consistent BD-DMA model and discuss its equivalent diagonal and ``beyond-diagonal'' forms. \textit{Second}, we propose in Sec.~\ref{sec_optimization} a general and efficient MC-aware optimization algorithm based on the diagonal formulation, applicable to any BD-DMA. \textit{Third}, we present in Sec.~\ref{sec_performance} results from physics-consistent simulations to demonstrate the performance enhancements with BD-DMAs compared to conventional DMAs in maximizing the channel gain for a single-input single-output (SISO) system. We systematically examine the role of the MC strength and the number of tunable vias.

\textit{Notation:} $\mathbf{A}_\mathcal{BC}$ denotes the block of the matrix $\mathbf{A}$ whose row [column] indices are in the set $\mathcal{B}$ [$\mathcal{C}$]. $\mathbf{a}_\mathcal{B}$ denotes the vector formed by the elements of $\mathbf{a}$ selected by $\mathcal{B}$. $\mathbf{a}^\top$ denotes the transpose of $\mathbf{a}$. $\mathbf{I}_a$ denotes the $a \times a$ identity matrix. $\mathbf{0}_{a,b}$ denotes the $a \times b$ zero matrix.

\section{System Model}
\label{sec_system_model}

\begin{figure}
    \centering
    \includegraphics[width=\columnwidth]{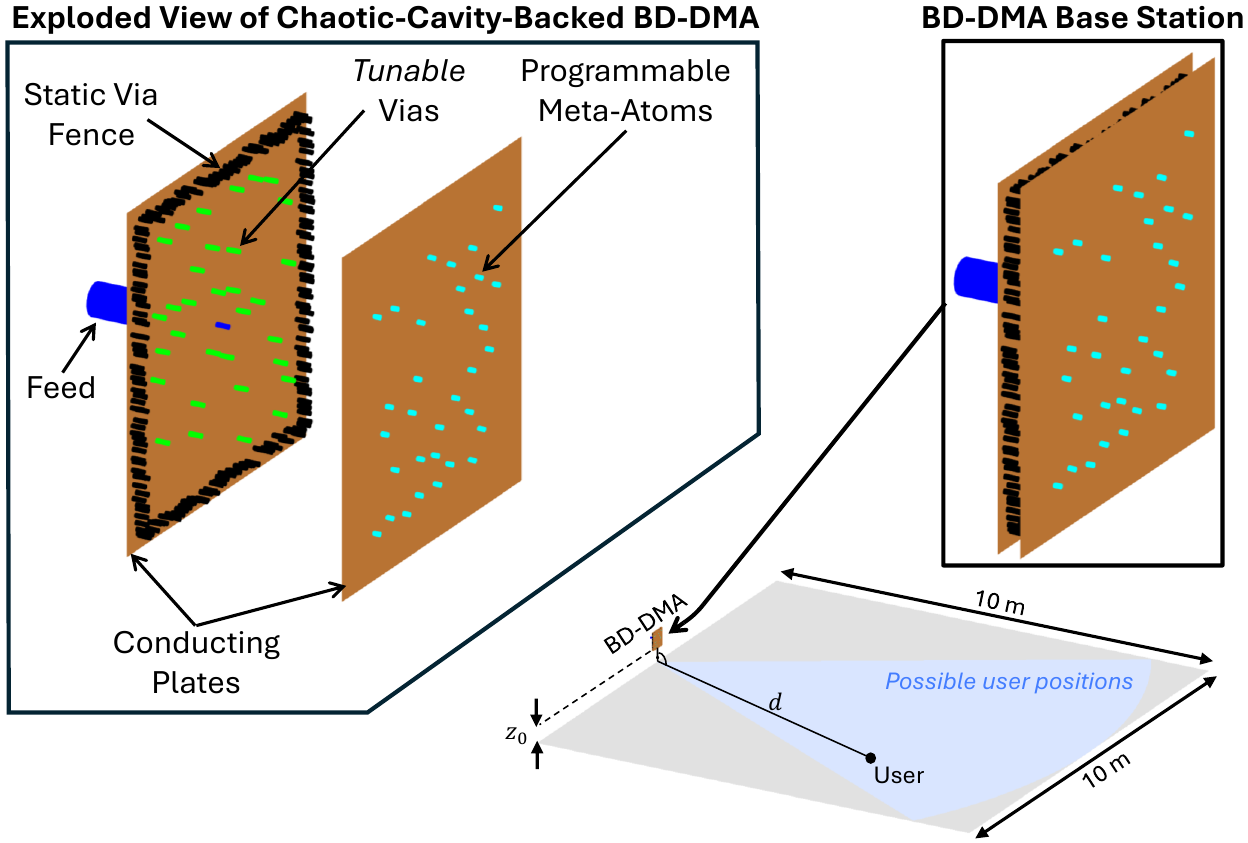}
    \caption{Exploded view of a single-feed chaotic-cavity-backed BD-DMA serving as base station. The green \textit{tunable} vias are the BD-DMA's distinguishing feature, enabling the reconfiguration of the coupling between the programmable meta-atoms. }
    \label{Fig1}
\end{figure}

In this section, we present a physics-consistent BD-DMA system model. 
For concreteness, we start by describing one possible BD-DMA hardware embodiment.
The starting point for this BD-DMA design is a conventional DMA design in which the \textit{static} intrinsic coupling between all $N_\mathrm{F}$ feeds and $N_\mathrm{M}$ meta-atoms is mediated by a cavity, defined by two parallel conducting plates and a fence of \textit{static} vias~\cite{yoo2018enhancing,sleasman2020implementation,prod2024mutual,prod2025benefits}. The feeds excite the cavity from the back; the programmable meta-atoms are patterned onto the front to controllably leak waves from the cavity, resulting in a programmable coherent wavefront.
Deviating from this conventional design, we add $N_\mathrm{V}$ \textit{tunable} vias at random locations inside the cavity.
The scattering properties of these vias can be tuned by (dis)connecting them from/to the upper conducting plate with PIN diodes; experimental implementations of this and alternative tuning mechanisms for vias  are surveyed in~\cite{entesari2015tunable}. Being part of the (otherwise static) cavity structure that mediates intrinsic coupling between all feeds and meta-atoms, these tunable vias hence enable \textit{reconfigurable} intrinsic coupling between all feeds and meta-atoms. Waves are scattered inside the DMA by feeds, meta-atoms and vias, but only meta-atoms radiate wave energy from inside the DMA to the external radio environment.
The single-feed chaotic-cavity-backed BD-DMA architecture that we consider for performance evaluations in Sec.~\ref{sec_performance} is shown in Fig.~\ref{Fig1}. We briefly comment on alternative BD-DMA embodiments in our outlook in Sec.~\ref{sec_conclusion}. The system model presented in this section applies to all BD-DMA architectures (i.e., it is not limited to the embodiment shown in Fig.~\ref{Fig1}).

\subsection{Physics-consistent BD-DMA model}

To formulate a physics-consistent model, we build on the established coupled-dipole model of conventional DMAs~\cite{yoo2019analytic,yoo2019analytic2,del2020learned,prod2025benefits} that was validated in commercial simulation software and experiments in~\cite{yoo2019analytic,yoo2019analytic2}.  
The BD-DMA features three types of primary entities: feeds, tunable vias, and programmable meta-atoms. Each of these entities is very small compared to the wavelength, justifying a dipole description. In total, we have $N=N_\mathrm{F}+N_\mathrm{V}+N_\mathrm{M}$ dipoles; the sets $\mathcal{F}$, $\mathcal{V}$ and $\mathcal{M}$ contain the indices of the dipoles associated with feeds, tunable vias, and programmable meta-atoms. The interaction of the $i$th dipole with an electromagnetic wave is characterized by its polarizability $\alpha_i$, which is fixed [reconfigurable] for dipoles associated with indices in $\mathcal{F}$ [$\mathcal{V}$ and $\mathcal{M}$]. The BD-DMA configuration can hence be defined by two vectors:
\label{eq_1}
\begin{equation}
    \mathbf{v} =  \left[  \alpha_i^{-1} \  \vert \  i \in \mathcal{V} \right] \in \mathbb{C}^{N_\mathrm{V}},\ \ 
    \mathbf{m} =  \left[  \alpha_i^{-1} \  \vert \  i \in \mathcal{M} \right] \in \mathbb{C}^{N_\mathrm{M}}.
\end{equation}
When an electromagnetic field $e_i$ is applied at the location and along the orientation of the $i$th dipole, it induces a dipole moment $p_i = \alpha_i e_i$. The $N$ dipoles are coupled to each other via \textit{background} Green's functions that capture the scattering properties of the static components of the considered DMA hardware, i.e., the chaotic cavity in Fig.~\ref{Fig1}. 
Specifically, $G_{ij}p_j$ is defined as the field created by a dipole moment $p_j$ at the position and along the orientation of the $j$th dipole that would be observed at the position and along the orientation of the $i$th dipole in the DMA  \textit{after removal of the entities modeled as dipoles}. 
We briefly discuss how to compute the values of $\alpha_i$ and $G_{ij}$ in Sec.~\ref{sec_performance}.

According to the superposition principle, we obtain a set of coupled equations, $e_i = \alpha_i^{-1} p_i =  e_i^\mathrm{inc} + \sum_{j=1}^N G_{ij} p_j$, that we can rewrite in matrix notation:
\begin{equation}
    \mathbf{A}\mathbf{p} = \mathbf{e}^\mathrm{inc} + \mathbf{G}\mathbf{p},
    \label{eq3}
\end{equation}
where $\mathbf{p} = [p_1,  \dots, p_N] \in \mathbb{C}^{N}$, $\mathbf{e}^\mathrm{inc} = [e_1^\mathrm{inc}, \dots, e_N^\mathrm{inc}] \in \mathbb{C}^{N}$, $\mathbf{A} = \mathrm{diag} \left( [\alpha_1^{-1},  \dots, \alpha^{-1}_N]  \right)\in \mathbb{C}^{N \times N}$, and the $(i,j)$th entry of $\mathbf{G}$ is $G_{ij}$. The input signal $\mathbf{x}\in\mathbb{C}^{N_\mathrm{F}}$ only excites the feeds, implying that  $\left[ e^\mathrm{inc}_i \  \vert \  i \in \mathcal{F} \right] \in \mathbb{C}^{N_\mathrm{F}}$ is proportional to $\mathbf{x}$ and $e_i^\mathrm{inc} = 0 \ \forall \ i \in \mathcal{V}\cup\mathcal{M}$. 
The self-consistent solution of~(\ref{eq3}) is 
\begin{equation}
    \mathbf{p} = \mathbf{W}^{-1}\mathbf{e}^\mathrm{inc},
    \label{eq_4}
\end{equation}
where $\mathbf{W}=\mathbf{A}-\mathbf{G}\in \mathbb{C}^{N \times N}$ is the system's interaction matrix.
To highlight the influence of the BD-DMA configuration defined in~(\ref{eq_1}), we partition $\mathbf{W}$ as follows:
\begin{equation}
    \mathbf{W}  = \begin{bmatrix} 
	\mathbf{W}^0_{\mathcal{FF}} & \mathbf{W}^0_{\mathcal{FV}} & \mathbf{W}^0_{\mathcal{FM}} \\
	\mathbf{W}^0_{\mathcal{VF}} & \mathbf{W}^0_{\mathcal{VV}} + \mathrm{diag}(\mathbf{v})& \mathbf{W}^0_{\mathcal{VM}}\\
    	\mathbf{W}^0_{\mathcal{MF}} & \mathbf{W}^0_{\mathcal{MV}}& \mathbf{W}^0_{\mathcal{MM}} + \mathrm{diag}(\mathbf{m})\\
	\end{bmatrix},
    \label{eq_5}
\end{equation}
where 
\begin{equation}
W^0_{ij}=\begin{cases}
\alpha_{i}^{-1} - G_{ii}, & i=j \in \mathcal{F}\\
 - G_{ij}, & \mathrm{otherwise}
\end{cases}.
\label{eq_6}
\end{equation}
All entries of $\mathbf{W}^0$ are static (they do not depend on $\mathbf{v}$ or $\mathbf{m}$).

Having evaluated the meta-atoms' dipole moments $\mathbf{p}_\mathcal{M}$ via~(\ref{eq_4}), the $k$-component (where $k\in\{x,y,z\}$) of the DMA's radiation pattern at some user position $\mathbf{r}$ is
\begin{equation}
\begin{split}
    e^{\mathrm{rad},k}  = \left( \mathbf{g}_k(\mathbf{r})\right)^\top \mathbf{p}_\mathcal{M} = \left( \mathbf{g}_k(\mathbf{r})\right)^\top \left[ \left(\mathbf{W} (\mathbf{v},\mathbf{m}) \right)^{-1} \right]_\mathcal{MF} \mathbf{e}^\mathrm{inc}_\mathcal{F},
    \end{split}
    \label{eq_7}
\end{equation}
where $\mathbf{r}_{\mathcal{M}_i}$ is the location of the $i$th meta-atom and $\mathbf{g}_k(\mathbf{r}) = \left[G_k^\mathrm{FS}(\mathbf{r},\mathbf{r}_i)  \  \vert \  i \in \mathcal{M} \right] \in \mathbb{C}^{N_\mathrm{M}}$ with $G_k^\mathrm{FS}(\mathbf{r},\mathbf{r}_{\mathcal{M}_i}) $ being the projection of the \textit{free-space} dyadic Green's tensor between positions $\mathbf{r}_{\mathcal{M}_i}$ and $\mathbf{r}$ onto the orientation of the field component $k$ at position $\mathbf{r}$ and onto the orientation of the $i$th dipole at position $\mathbf{r}_{\mathcal{M}_i}$; (9.18) in~\cite{jackson1999electrodynamics} provides the closed-form expression for $G_k^\mathrm{FS}(\mathbf{r},\mathbf{r}_{\mathcal{M}_i})$. 

In line with theoretical studies on near-field applications of conventional DMAs~\cite{zhang2022beam,azarbahram2024waveform,zhang2025near}, we consider in Sec.~\ref{sec_performance} a static free-space scenario with a single-feed BD-DMA base station and a non-invasive receiver.
The end-to-end channel from the BD-DMA port to the receiver port is thus $e^{\mathrm{rad},k}$ for a receiver capturing the radiated field's $k$-component.

\subsection{``Beyond-diagonal'' programmability revealed in reduced representation}

\subsubsection{Diagonal representation}
The BD-DMA's programmability appears as diagonal matrix in (\ref{eq_5}), as expected for \textit{any} wave system parametrized by tunable lumped elements~\cite{del2025physics}. To see this diagonal form more clearly, we choose a partition of the BD-DMA's primary entities into static ones ($\mathcal{F}$) and reconfigurable ones ($\mathcal{R}=\mathcal{V}\cup\mathcal{M}$):
\begin{equation}
    \mathbf{W}  = \begin{bmatrix} 
	\mathbf{W}^0_{\mathcal{FF}} & \mathbf{W}^0_{\mathcal{FR}} \\
	\mathbf{W}^0_{\mathcal{RF}} & \mathbf{W}^0_{\mathcal{RR}} + \mathrm{diag}(\mathbf{c})\\
	\end{bmatrix},
    \label{eq_8}
\end{equation}
where $\mathbf{c}=[\mathbf{v}^\top,\ \mathbf{m}^\top]^\top\in \mathbb{C}^{N_\mathrm{V}+N_\mathrm{M}}$. Inserting (\ref{eq_8}) into (\ref{eq_4}) and applying the block matrix inversion lemma, we obtain
\begin{equation}
    \mathbf{p}_\mathcal{M} = - \mathbf{U} \left[ \tilde{\mathbf{W}}+\mathrm{diag}(\mathbf{c})\right]^{-1}\mathbf{W}^0_\mathcal{RF}\left(\mathbf{W}^0_\mathcal{FF}\right)^{-1}\mathbf{e}^\mathrm{inc}_\mathcal{F},
    \label{eq_9}
\end{equation}
where $\mathbf{U} = \begin{bmatrix}
\mathbf{0}_{N_{\mathrm{M}},N_{\mathrm{V}}} & \mathbf{I}_{N_{\mathrm{M}}}
\end{bmatrix}$ and
\begin{equation}
        \tilde{\mathbf{W}} = \mathbf{W}^0_\mathcal{RR} - \mathbf{W}^0_\mathcal{RF} \left( \mathbf{W}^0_\mathcal{FF} \right)^{-1} \mathbf{W}^0_\mathcal{FR}.
\end{equation}
The diagonal representation of the BD-DMA's programmability is now evident in (\ref{eq_9}), where all terms are static except for $\mathrm{diag}(\mathbf{c})$. We can insert (\ref{eq_9}) into (\ref{eq_7}) to obtain the DMA's radiation pattern.

\subsubsection{Beyond-diagonal representation}
Alternatively, we can consider a representation reduced to the two types of primary entities that also appear in a conventional DMA ($\mathcal{P} = \mathcal{F}\cup\mathcal{M}$) to facilitate comparisons. In the reduced representation, the tunable vias are treated as part of the ``background'' scattering in the cavity structure that mediates intrinsic MC between feeds and meta-atoms. 
Therefore, the role of the tunable vias as reconfigurable intrinsic coupling mechanism is apparent in the reduced-basis representation. Similar reduced-basis representations were recently used to reveal hidden symmetries in metamaterials~\cite{HiddenSymmetry} and to efficiently evaluate RIS-parametrized channels~\cite{prod2023efficient}.
We denote reduced-basis variables with a ring ( $\mathring{}$ ). 
Analogous to Sec.~III in~\cite{prod2023efficient}, we obtain a reduced-basis representation with the following reduced interaction matrix:
\begin{equation}
    \mathring{\mathbf{W}}  = \begin{bmatrix} 
	\mathring{\mathbf{W}}^0_{\mathcal{FF}} & \mathring{\mathbf{W}}^0_{\mathcal{FM}} \\
	\mathring{\mathbf{W}}^0_{\mathcal{MF}} & \mathring{\mathbf{W}}^0_{\mathcal{MM}} + \mathrm{diag}(\mathbf{m})\\
	\end{bmatrix}\in \mathbb{C}^{N_2\times N_2},
    \label{eq_A}
\end{equation}
where $N_2 = N_\mathrm{F}+N_\mathrm{M}$ and 
\begin{equation}
    \mathring{\mathbf{W}}^0 = \mathbf{W}_{\mathcal{PP}}^0 - \mathbf{W}_{\mathcal{PV}}^0 \left(\mathbf{W}_{\mathcal{VV}}^0+\mathrm{diag}(\mathbf{v})\right)^{-1} \mathbf{W}_{\mathcal{VP}}^0.
    \label{eq_12}
\end{equation}
All entries of the reduced-basis $\mathring{\mathbf{W}}^0$ defined in (\ref{eq_12}) generally depend on $\mathbf{v}$ and are hence reconfigurable, in contrast to $\mathbf{W}^0$ defined in (\ref{eq_6}) whose entries are static. The reduced-basis representation yields an alternative expression for $\mathbf{p}_\mathcal{M}$:
\begin{equation}
    \mathbf{p}_\mathcal{M} = - \left[ \mathring{\tilde{\mathbf{W}}}+\mathrm{diag}(\mathbf{m})\right]^{-1}\mathring{\mathbf{W}}^0_\mathcal{MF}\left(\mathring{\mathbf{W}}^0_\mathcal{FF}\right)^{-1}\mathbf{e}^\mathrm{inc}_\mathcal{F},
        \label{eq_13}
\end{equation}
where 
\begin{equation}
        \mathring{\tilde{\mathbf{W}}} = \mathring{\mathbf{W}}^0_\mathcal{MM} - \mathring{\mathbf{W}}^0_\mathcal{MF} \left( \mathring{\mathbf{W}}^0_\mathcal{FF} \right)^{-1} \mathring{\mathbf{W}}^0_\mathcal{FM}.
\end{equation}

\subsubsection{Comparison}
Clearly, (\ref{eq_9}) and (\ref{eq_13}) must be equivalent since both are physics-consistently derived for the same physical system. Whereas $\mathrm{diag}(\mathbf{c})$ is the only reconfigurable term in (\ref{eq_9}), both $\mathring{\tilde{\mathbf{W}}}$ and $\mathrm{diag}(\mathbf{m})$, as well as $\mathring{\mathbf{W}}^0_\mathcal{MF}$ and $\mathring{\mathbf{W}}^0_\mathcal{FF}$, are reconfigurable in (\ref{eq_13}). The entries of $\mathrm{diag}(\mathbf{c})$ or $\mathrm{diag}(\mathbf{m})$ can be chosen independently. In contrast, the entries of $\mathring{\tilde{\mathbf{W}}}$, $\mathring{\mathbf{W}}^0_\mathcal{MF}$ and $\mathring{\mathbf{W}}^0_\mathcal{FF}$ cannot be chosen independently and arbitrarily, being parametrized by $\mathrm{diag}(\mathbf{v})$ according to the mathematical structure of (\ref{eq_12}). Analogously, the entries of the ``beyond-diagonal'' matrix parameterizing a BD-RIS cannot be chosen independently and arbitrarily because they depend on a diagonal programmable matrix via a complicated relation akin to (\ref{eq_12}), see (5) in~\cite{del2025physics}. However, theoretical studies on BD-RIS~\cite{shen2021modeling,li2024beyond} usually assume for simplicity that the entries of the beyond-diagonal matrix can be optimized independently from each other up to some global constraints like energy conservation. We do not make this simplifying assumption in this Letter but it could be explored in future work to obtain upper bounds on BD-DMA performance. 

In the limit in which the reconfigurability of the tunable vias vanishes, i.e., when $\mathbf{v}$ is fixed, the BD-DMA specializes to the conventional DMA because the intrinsic MC between all feeds and meta-atoms becomes static. With static $\mathring{\tilde{\mathbf{W}}}$, $\mathring{\mathbf{W}}^0_\mathcal{MF}$ and $\mathring{\mathbf{W}}^0_\mathcal{FF}$ in that case, (\ref{eq_13}) resembles the physics-consistent model of a conventional DMA~\cite{yoo2019analytic,yoo2019analytic2,del2020learned,prod2025benefits}.

\section{Mutual-Coupling-Aware Optimization}
\label{sec_optimization}

Our discussion in the previous section indicates that the BD-DMA system model based on the diagonal representation in (\ref{eq_9}) is preferable for optimization because all reconfigurable variables therein can be controlled independently. In contrast, the reconfigurable variables in the equivalent ``beyond-diagonal'' representation based on (\ref{eq_13}) cannot be controlled independently. Moreover, the mathematical structure of (\ref{eq_9}) is the same as for a conventional DMA, such that existing MC-aware optimization algorithms for conventional DMAs (e.g., gradient descent~\cite{del2020learned,qian2022noise}, coordinate descent~\cite{prod2025benefits}) can be directly applied to BD-DMAs. Based on the diagonal representation in (\ref{eq_9}), all tunable lumped elements are treated the same in $\mathbf{c}$, irrespective of whether they reconfigure local meta-atom properties or intrinsic MC between meta-atoms.

\begin{algorithm}[b]
\footnotesize
\label{Alg_CoordAsc}
Evaluate costs $\left\{\mathcal{C}\right\}_{\mathrm{init}}$ for $512$ random BD-DMA configurations $\left\{\mathbf{c}\right\}_{\mathrm{init}}$.\\
Select lowest-cost configuration $\mathbf{c}_{\mathrm{curr}}$ from $\left\{\mathbf{c}\right\}_{\mathrm{init}}$ such that $\mathcal{C}_{\mathrm{curr}} = \mathcal{C}\left(\mathbf{c}_{\mathrm{curr}}\right) = \mathrm{min}\left(\left\{\mathcal{C}\right\}_{\mathrm{init}}\right)$.\\
$k \gets 0$, $i \gets 0$ \\
\While{$k < N_\mathrm{V}+N_\mathrm{M}$}{
    $\mathbf{c}_\mathrm{temp} \gets \mathbf{c}_\mathrm{curr}$  with  $(1+\mathrm{mod}(i,N_\mathrm{V}\!+\!N_\mathrm{M}))$th entry flipped.\\
    $\mathcal{C}_{\rm temp} \gets \mathcal{C}\left(\mathbf{c}_\mathrm{temp}\right)$.\\
    \eIf{
    $\mathcal{C}_{\rm temp} < \mathcal{C}_{\rm curr}$
    } {
    $\mathbf{c}_\mathrm{curr} \gets \mathbf{c}_\mathrm{temp}$ \\
    $\mathcal{C}_\mathrm{curr} \gets \mathcal{C}_\mathrm{temp}$ \\
    $k \gets 0$
    } {$k \gets k+1$}
    $i \gets i+1$\\
    }
\KwOut{$\mathbf{c}_{\rm curr}$ and $\mathcal{C}_\mathrm{curr}$.}
\caption{Binary coordinate descent optimization.}
\end{algorithm}

Most existing DMA prototypes are constrained to 1-bit programmable lumped elements to keep the circuitry that controls the lumped elements' bias voltages simple~\cite{sleasman2015dynamic,sleasman2020implementation,prod2025benefits}.
Thus, we consider binary programmability constraints in this Letter, i.e., for each entry of $\mathbf{c}$ there are only  two possible values.
Although special gradient-descent algorithms can accommodate such binary constraints~\cite{del2020learned,qian2022noise}, we opt for the coordinate descent algorithm summarized in Algorithm~\ref{Alg_CoordAsc} which is  naturally adapted to binary constraints and can leverage fast forward evaluations of the system model based on the Woodbury identity~\cite{prod2023efficient}.
Initially, Algorithm~\ref{Alg_CoordAsc} establishes a dictionary of 512 pairs $\{\mathbf{c},\mathcal{C}(\mathbf{c})\}$ of random BD-DMA configurations and their associated costs. The lowest-cost dictionary entry serves as initialization for a coordinate descent optimization that repeatedly loops over all $N_\mathrm{V}+N_\mathrm{M}$ reconfigurable entities, testing for one entity at a time whether flipping its binary configuration reduces the cost. 
The repeated loops are 
required because the cost depends non-linearly on $\mathbf{c}$.\footnote{Due to MC, the mapping from $\mathbf{c}$ to $\mathcal{C}(\mathbf{c})$ is non-linear irrespective of the cost's dependence on the BD-DMA's radiated field.} 
Algorithm~\ref{Alg_CoordAsc} runs until it has completed an entire loop 
without updating $\mathbf{c}$, which indicates convergence.
The generic nature of Algorithm~\ref{Alg_CoordAsc} implies that it can be applied to any BD-DMA architecture and any definition of the cost.

\section{Performance Evaluation}
\label{sec_performance}

In this section, we analyze the performance of the proposed BD-DMA as a 10~GHz base station in free space. We aim to maximize the channel gain $\beta=\sum_k \left| e^{\mathrm{rad},k} \right|^2$ to a single user equipment (UE), assuming the UE exploits both non-zero field components (which carry the same signal) based on non-invasive receivers. Hence, we use Algorithm~\ref{Alg_CoordAsc} with $\mathcal{C}=-\beta$. The single-user channel gain is a commonly considered key performance indicator (see, e.g.,~\cite{li2024beyond}) because it directly correlates with end-to-end communication metrics (e.g., capacity, bit error rate, or throughput). Explorations of multi-user scenarios are deferred to future work because of their complexity and specificity to choices like the noise level.

The UE's allowed positions are restricted to the wedge highlighted in blue in Fig.~\ref{Fig1}, excluding locations at grazing angles with respect to the BD-DMA's surface.  
Its feed is located at its center which is placed at a height $z_0$ relative to the plane $\Pi$ of allowed UE positions. $d$ denotes the shortest distance between the UE and the normal to $\Pi$ running through the BD-DMA's center. 
The BD-DMA's footprint is $40\times 40\ \mathrm{cm}^2$ but all $N_\mathrm{M}=32$ programmable meta-atoms are located within a $30\times 30\ \mathrm{cm}^2$ area. The static via fence consists of 200 vias and is deliberately irregular to induce wave chaos. 
Tunable vias and programmable meta-atoms are randomly located within the cavity, up to a minimum-separation constraint.

We analytically evaluate the background Green's functions between our BD-DMA's $N$ primary entities similar to~\cite{yoo2019analytic,yoo2019analytic2}.\footnote{Specifically, we initially treat the static vias as dipoles, too, and then change to a representation reduced to our primary entities, similar to Sec.~III in~\cite{prod2023efficient}, yielding the background Green's functions used in Sec.~\ref{sec_system_model}.} We account for losses in the substrate between the two conducting plates with a complex substrate wavenumber $k_0(1-\jmath\gamma)$, where $k_0$ is the wavenumber in the lossless case and $\gamma$ is the lossiness factor~\cite{prod2025benefits}. By changing the value of the substrate lossiness, we access different regimes of MC strength~\cite{prod2025benefits}. We choose the dipole polarizability values based on those extracted from full-wave simulations for a complementary electric inductive-capacitive (cELC) meta-atom embedded in a waveguide structure in~\cite{pulido2017polarizability}. The same procedure can be used to establish a dictionary mapping the reconfigurable via states to the corresponding polarizabilities; however,  we assume for simplicity that the two possible values of $\alpha_i\in\mathcal{V}$ are 1~\% and 99~\% of the static via polarizability which is known analytically~\cite{yoo2019analytic}.
The meta-atoms in our BD-DMA are magnetic dipoles oriented perpendicular to $\Pi$.

\begin{figure}
    \centering
    \includegraphics[width=\columnwidth]{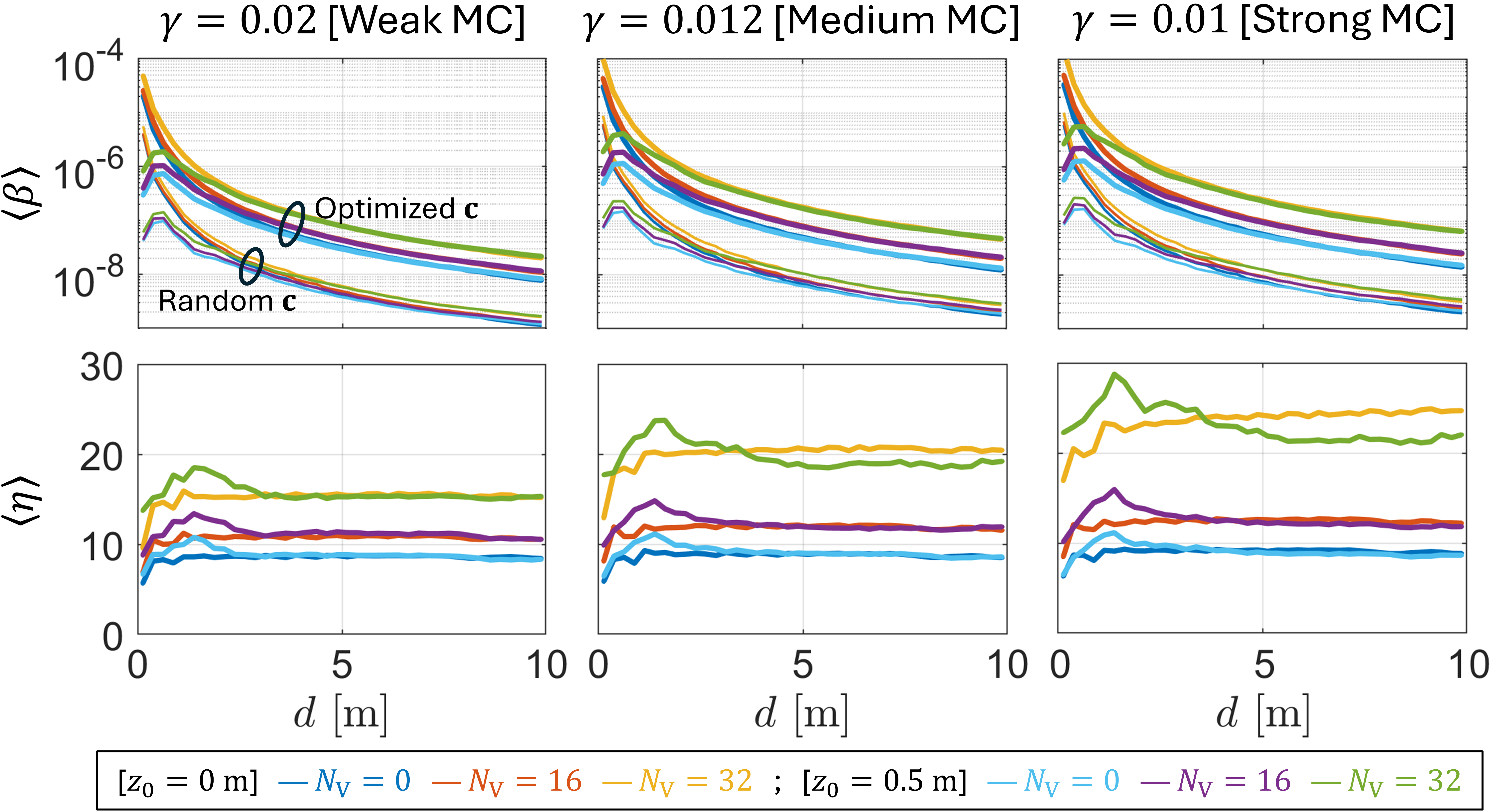}
    \caption{Top row: Average channel gain $\langle \beta \rangle$ with random and optimized BD-DMA configurations. Bottom row: Average channel gain enhancement $\langle \eta \rangle$. Results are shown for three different MC strengths (see subfigure titles), three different values of $N_\mathrm{V}$ ($N_\mathrm{V}=0$ corresponds to a conventional DMA with fixed intrinsic MC) and two values of $z_0$ (see legend). }
    \label{Fig2}
\end{figure}

In Fig.~\ref{Fig2}, we systematically compare the average channel gain enhancement for three different levels of MC strength (by changing the substrate lossiness: $\gamma = 0.02,\ 0.012,\ 0.01$), three different values of $N_\mathrm{V}$ (0, 16, and 32) and two values of $z_0$ (0~m and 0.5~m). For each combination of these parameters, we applied Algorithm~\ref{Alg_CoordAsc} independently to 6402 grid points inside the blue wedge in Fig.~\ref{Fig1}. For $N_\mathrm{V}=0$, the BD-DMA is simply a conventional DMA with static intrinsic couplings between meta-atoms, serving as benchmark. The channel gain enhancement enabled by reconfigurable intrinsic coupling mechanisms is seen to scale super-linearly with $N_\mathrm{V}$ (the gap from $N_\mathrm{V}=16$ to 32 is much larger than from $N_\mathrm{V}=0$ to 16) and to increase with the MC strength. The latter makes sense: the stronger the MC between meta-atoms, the more benefit can be derived from being able to reconfigure it. We also observe markedly different behaviors at small values of $d$ for different BD-DMA heights $z_0$. 
As seen in the top row of Fig.~\ref{Fig2}, the average channel gain decays monotonically with $d$ for $z_0=0\ \mathrm{m}$ but peaks at a small value of $d$ for  $z_0=0.5\ \mathrm{m}$ because of the non-isotropic radiation pattern of the vertically oriented meta-atoms. 
As expected, for larger $d$, the differences in $z_0$ become negligible and the associated curves converge.

\section{Conclusion}
\label{sec_conclusion}

To summarize, we introduced a novel DMA architecture coined BD-DMA that distinguishes itself from conventional DMAs by its reconfigurable (as opposed to fixed) intrinsic coupling mechanisms between meta-atoms. We derived a physics-consistent system model with two equivalent representations: one with uncorrelated, ``diagonal'' programmability and one in a reduced basis with correlated, ``beyond-diagonal'' programmability. Building on the former, we proposed an efficient and general MC-aware optimization algorithm. In physics-consistent simulations we observed that the BD-DMA benefits grow with the intrinsic MC strength and scale super-linearly with the number of tunable elements dedicated to reconfiguring the intrinsic coupling.

Looking forward, alternative implementations and topologies of reconfigurable intrinsic coupling mechanisms between meta-atoms can be explored. For instance, tunable circuital coupling networks can be embedded into lower PCB layers, and hence they can also be envisioned to upgrade the widely studied stacked-microstrip-based DMAs to group-connected or fully-connected BD-DMAs.

\bibliographystyle{IEEEtran}
%\bibliography{references2}

\providecommand{\noopsort}[1]{}\providecommand{\singleletter}[1]{#1}%

\end{document}